\documentclass[onecolumn,secnumarabic,amssymb, aps, prfluids,amsmath]{revtex4}

\usepackage{graphicx}
\usepackage{natbib}
\usepackage{amsmath}
\usepackage{color}

\newcommand{\Pe}{\mbox{Pe}}

\begin{document}

\title{Bouncing, chasing or pausing: asymmetric collisions of active droplets}%

\author{Kevin Lippera}
\affiliation{LadHyX -- D\'epartement de M\'ecanique, CNRS -- Ecole Polytechnique, Institut Polytechnique de Paris, 91128 Palaiseau, France}
\author{Michael Benzaquen}
\affiliation{LadHyX -- D\'epartement de M\'ecanique, CNRS -- Ecole Polytechnique, Institut Polytechnique de Paris, 91128 Palaiseau, France}
\author{S\'ebastien Michelin}
\email{sebastien.michelin@ladhyx.polytechnique.fr}
\affiliation{LadHyX -- D\'epartement de M\'ecanique, CNRS -- Ecole Polytechnique, Institut Polytechnique de Paris, 91128 Palaiseau, France}
\date{\today}

\begin{abstract}

Chemically-active droplets exhibit complex avoiding trajectories. While heterogeneity is inevitable in active matter experiments, it is mostly overlooked in their modelling. Exploiting its geometric simplicity, we fully-resolve the head-on collision of two swimming droplets of different radii and demonstrate that even a small contrast in size critically conditions their collision and subsequent dynamics. We identify three fundamentally-different regimes. The resulting high sensitivity of pairwise collisions is expected to profoundly affect their collective dynamics.
\end{abstract}

\maketitle

Micron-size droplets suspended in a surfactant solution may display the basic features of active systems, as a result of complex interplays of thermodynamic and interfacial processes coupled to hydrodynamic flows. Such systems typically convert chemical free energy to spontaneously develop complex deformations~\cite{pimienta2011,Caschera13}, unsteady dynamics~\cite{herminghaus2014interfacial,wodlei2018} and/or self-propulsion~\cite{thutupalli2011swarming,izri2014self,Maass16}. For this very reason, they have recently received much attention from the physics community as canonical active matter systems to analyse self-propulsion, collective dynamics and self-organization at the micron scale~\cite{bechinger2016,zottl2016}. 

Of major interest to physicists is their potential ability to reproduce the complex behaviour of biological swimmers without resorting to intrinsically-biological traits but simply as a  result of hydrodynamic or chemical interactions. In contrast with other synthetic micro-systems, {active droplets} do not rely on any external magnetic~\cite{ghosh2009} or mechanical forcing~\cite{wang2012}, nor any intrinsic broken symmetry~\cite{Moran17,kummel2013}. Instead, the emergence of spontaneous dynamics stems from the nonlinear coupling of physico-chemical processes and convective transport. Beyond steady and directed self-propulsion, active droplets may exhibit complex curling trajectories~\cite{Kruger16,Suga18,hokmabad2019}, random diffusive behaviour~\cite{Suga18,izzet2019} and unsteady self-propulsion~\cite{thutupalli2013} at the individual level, and collective dynamics such as avoidance~\cite{Jin17}, aggregation~\cite{seemann2016,Kruger16b}, polar alignment~\cite{thutupalli2011swarming}, crystallization~\cite{thutupalli2018flow} and train formation~\cite{Illien19one}. 

Such behaviour diversity is partly the reflection of a large variety of physical systems, and in particular of the origin of physico-chemical activity. Self-propelled droplets can be broadly sorted into two main categories, depending on whether they use chemical reactions or phase solubilization~\cite{herminghaus2014interfacial,Maass16,seemann2016}. In both cases, the emergence of self-propulsion results from non-homogenous interfacial tension (or Marangoni stresses), generated by a non-uniform distribution of chemical species in the immediate vicinity of their surface. 

Dissolving droplets -- which are the focus of the present Letter -- exploit the slow solubilization of their fluid content into the surrounding surfactant-rich phase, either by forming new filled micelles (molecular route~\cite{moerman2017solute}) or by direct transfer to existing micelles (micellar route~\cite{izri2014self}). Both routes involve large molecular compounds or solutes, whose interaction with the interface directly affects the local surface tension that drives fluid flows and locomotion, thereby providing the droplet with the activity and mobility required for self-propulsion~\cite{herminghaus2014interfacial}. 
 \begin{figure}
\centering
\includegraphics[width=.5\textwidth]{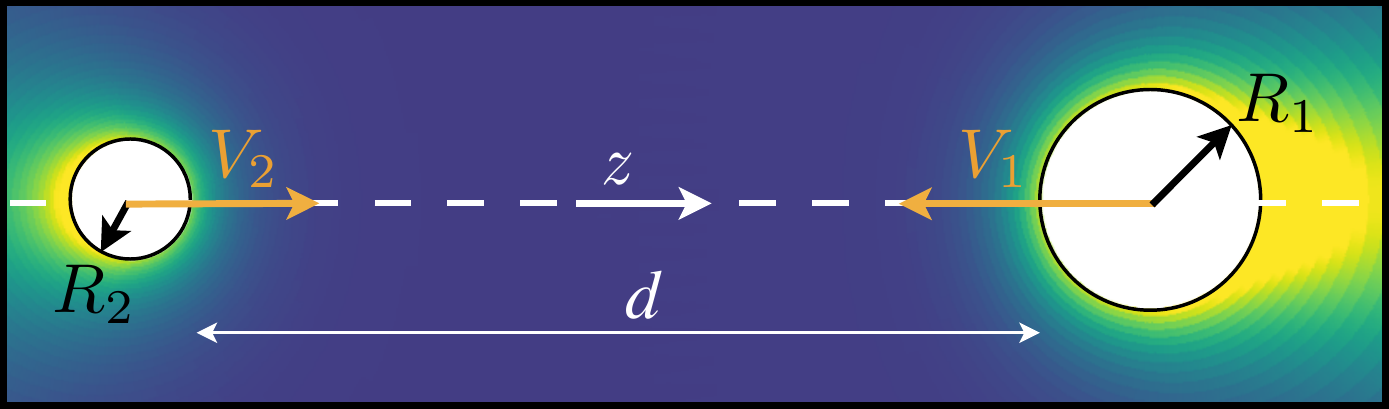}
\caption{Collision of two chemically-active self-propelled droplets. The color shades show typical solute concentration distributions around the moving droplets. \vspace{-.5cm}}
\label{PhysicalScheme}
\end{figure}

In contrast with intrinsically-asymmetric Janus colloids~\cite{Moran17}, active droplets spontaneously break a directional symmetry as a result of an instability stemming from  the non-linear coupling of the solutes' convective transport to the self-generated Marangoni flows~\cite{michelin2013spontaneous,izri2014self,Maass16}. Recent modelling efforts have also uncovered several important features of individual self-propulsion~\cite{schmitt2013,yoshinaga2014,morozov2019self,morozov2019nonlinear}. In contrast, detailed modelling of the interaction of two or multiple droplets has so far been rather elusive, due to the non-linearity of the advective problem, and as a result, most of the existing models rely on far-field approximations for which superposition principles can be adapted~\cite{yabunaka2016collision,moerman2017solute}. Yet, most experiments report much denser situations; {the present Letter thus focuses on such near-field interactions, which} are likely critical to the emergence of collective behaviour.

Exploiting the simplicity of a two-sphere geometry, we were recently able to fully-resolve the symmetric collision of two droplets~\cite{Lippera19collisions}, and provided a critical insight on how the nonlinear convective transport conditions the collision. To note, controlling experimentally the exact droplet radius, which slowly varies over time~\cite{izri2014self}, is however impossible, giving rise to heterogeneous systems and non-symmetric in practice. \\

In this Letter, using a fully-coupled model of the detailed solute and flow dynamics, we demonstrate how even a small contrast in the droplets' size critically {conditions} their head-on collision and subsequent dynamics, thus likely affecting substantially the complex collective behavior of active droplets suspensions.
To this end, we consider the axisymmetric dynamics of two droplets of fixed radii $R_1$ and $R_2$, and of identical chemical properties, with a minimum surface-to-surface distance $d$ (Figure~\ref{PhysicalScheme}). The chemical activity of the droplet is generically modelled here as a fixed-rate surface release of solute -- coined $\mathcal{A}$ --  into the outer fluid. The interaction of this solute with the droplets' interface modifies its effective local surface tension, introducing a chemically-induced Marangoni stress~\cite{anderson1989colloid}. At the droplets' surface, activity and mobility therefore impose 
\begin{align}
D\mathbf{n}\cdot\nabla c=-\mathcal{A},\qquad\qquad \mathbf{n}^\perp\cdot[\boldsymbol\sigma-\boldsymbol{\tilde\sigma}]\cdot\mathbf{n}=-{\gamma_c}\nabla_\parallel c\ ,\label{eq:bcs}
\end{align}
where $\mathbf{n}$ and $\mathbf{n}^\perp$ are the outward normal vector and surface projection operator along the interface, respectively{.} $c$ {denotes} the solute concentration and $D$ its molecular diffusivity{.} {In addition for} small concentration variations, {the solute-induced change in surface tension, $\gamma_c=(\partial \gamma/\partial c)$}, is constant, and $\boldsymbol\sigma$ and $\boldsymbol{\tilde\sigma}$ are respectively the Cauchy stress tensors of the outer and inner Newtonian fluids of viscosities $\eta$ and $\tilde\eta$. Droplets must be negatively auto-chemotactic in order to self-propel~\cite{michelin2013spontaneous,Jin17}{: they must be repelled by the solute they release, which is advected into their wake by the surface Marangoni flows.}  This, in turn, imposes {$\mathcal{A}\gamma_c>0$}; both are taken positive in the following.

  Chemical solutes involved in the spontaneous emulsification of active droplets are large molecular compounds~\cite{herminghaus2014interfacial} and  diffuse slowly so that their concentration~$c$ in the outer phase follows an advection-diffusion equation 
\begin{equation}
\frac{\partial c}{\partial t}+\mathbf{u}\cdot\nabla c=D\nabla^2 c\ .
\end{equation}
Due to the microscopic size of the droplets, inertial effects are negligible, and the velocity of the fluid is found by solving Stokes' equations in each phase, coupled through the Marangoni condition, Eq.~\eqref{eq:bcs}. Each droplet is hydrodynamically force-free, which {determines uniquely their velocities, $V_1$ and $V_2$,  measured positively along the axis of symmetry $z$.} 
The full nonlinear coupling of the Stokes and chemical transport problems is solved numerically for arbitrary distance using an efficient scheme based on a time-dependent bispherical grid presented in detail in Ref.~\cite{Lippera19collisions}. 

The problem is now re-scaled using $R_2$, $\mathcal{V}_2$ and {$\mathcal{A}R_2/D$} as reference length, velocity and concentration respectively, where ${\mathcal{V}_i=\mathcal{A}{\gamma_c} R_i/[D(2\eta+3\tilde\eta)]}$ is the {passive} drift velocity of a droplet of radius $R_i$ in an externally-imposed gradient $\mathcal{A}/D$~\cite{anderson1989colloid}. In addition to the viscosity ratio $\eta/\tilde\eta$ (which is set to unity for simplicity here), the problem is fully-determined by two independent non-dimensional parameters, namely the specific P\'eclet number (or advection-diffusion ratio) $\Pe_i=R_i\mathcal{V}_i/D$  of each droplet. Together they are related to the size ratio $\xi=R_1/R_2=\sqrt{\Pe_1/\Pe_2}$.  The P\'eclet number plays a key role for an individual droplet's dynamics, and for $\Pe_i\geq \Pe_c=4$, a droplet of radius $R_i$ starts swimming spontaneously as a result of a transcritical bifurcation~\cite{izri2014self,morozov2019self}.  

{To identify} the effect of the size ratio, $\xi$, the results are reported in the following  in terms of $\xi\geq 1$ and of the P\'eclet number of the smaller droplet, $\Pe_2\geq 4$ for the time-dependent dynamics of two droplets initially located far apart ($d\gg 1$) and swimming toward each other ($V_2>0$ and $V_1<0$).  Depending on the value of $(\xi,\Pe_2)$, three dynamic collision regimes are observed, and are analyzed in  detail in the following.\\

\begin{figure}[t!]
\begin{center}
\begin{tabular}{cc}
\multicolumn{2}{c}{\includegraphics[width=.45\textwidth]{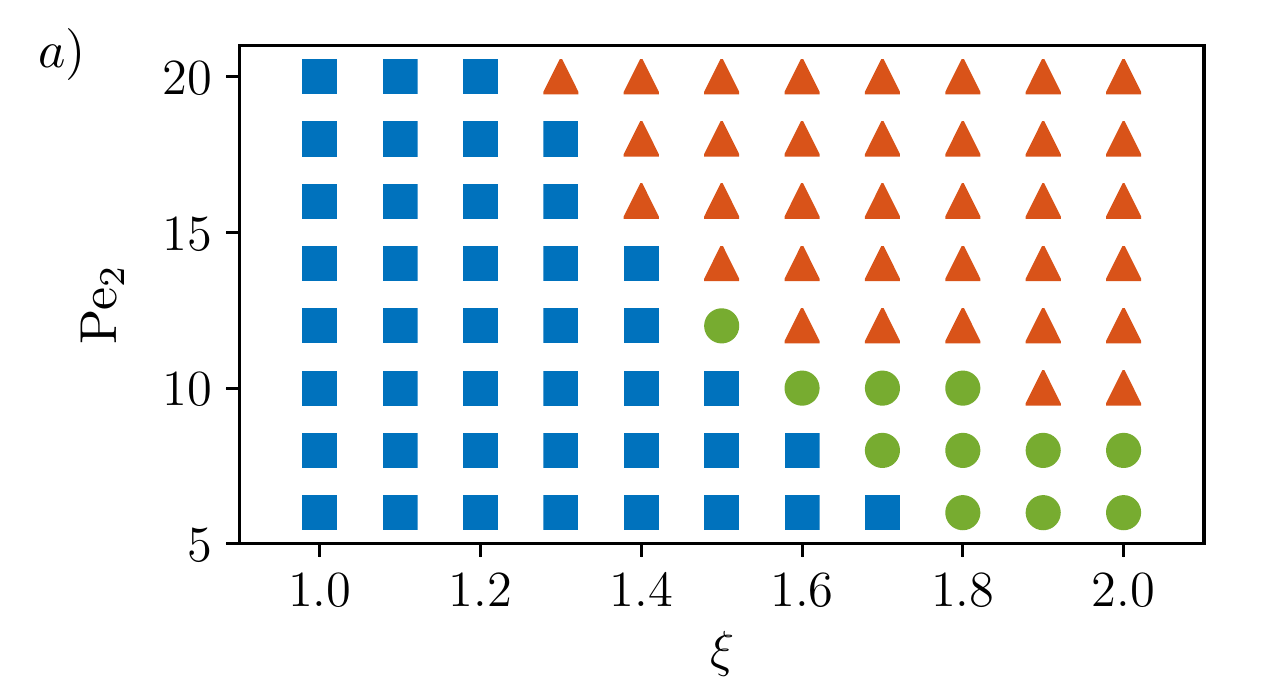}} \\
\includegraphics[width=.5\textwidth]{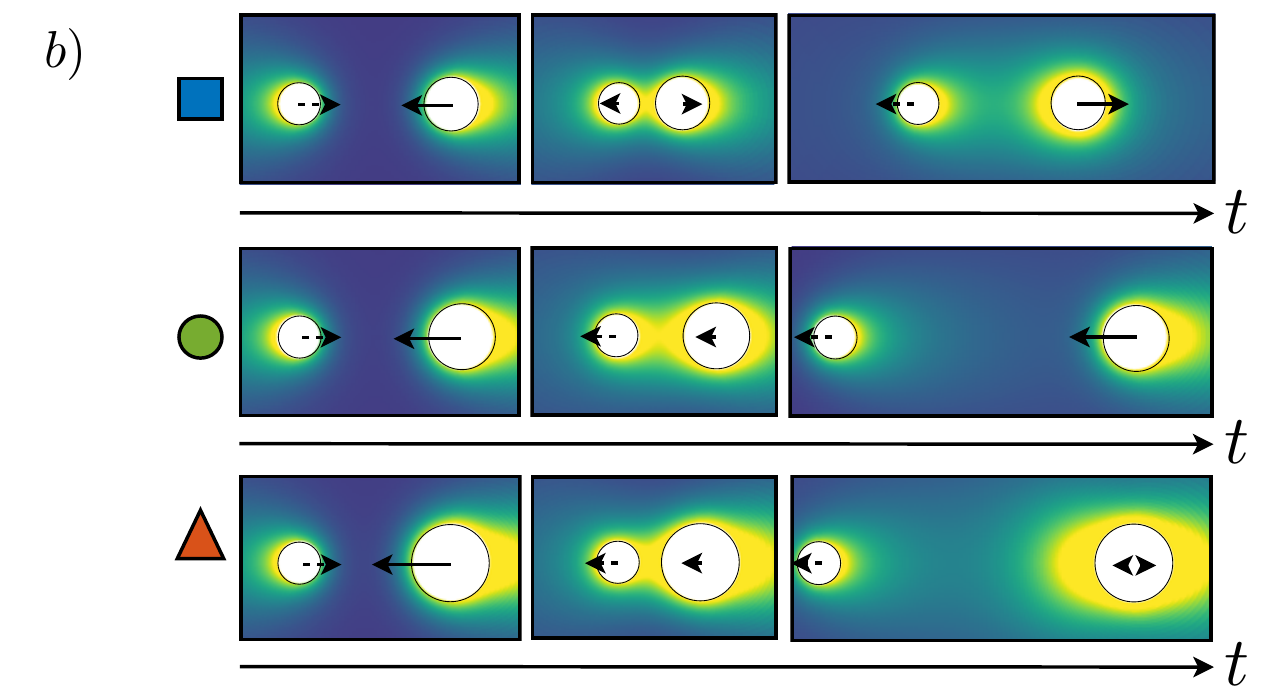} &
\includegraphics[width=.5\textwidth]{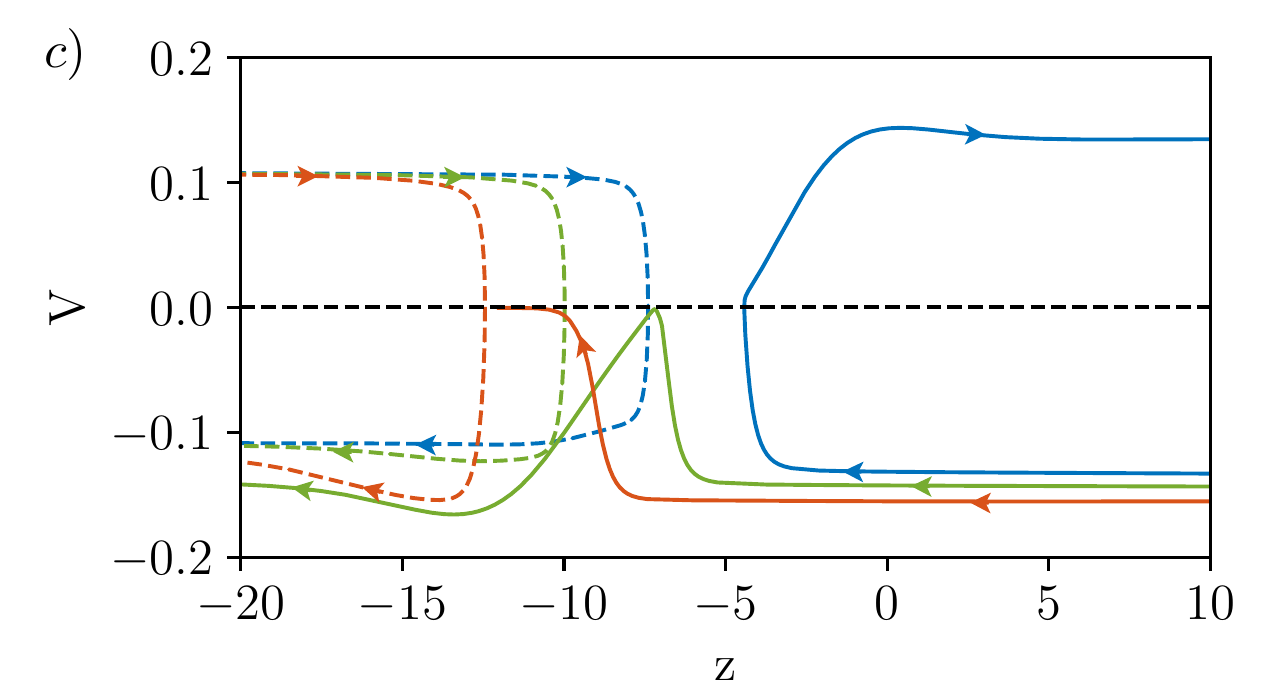}
\end{tabular}
\caption{(a) Regime selection for the asiymmetric dynamic collision of two droplets for varying $(\xi,\Pe_2)$: rebound (blue square), chasing (green disk) and pausing regimes (red, triangle). (b) Snapshots of the corresponding dynamics for each regime, showing the droplets' instantaneous velocity (arrow) and solute concentration distribution (color) (movies available in Supplemental Material~\cite{SI}). (c) Evolution of the small (dashed) and large (solid) droplets'  velocity with their position during three representative collision regimes  at $\Pe=12$: $\xi=1.3$ (blue, rebound), $\xi=1.5$ (green, chasing) and $\xi=1.8$ (red, pausing).  \vspace{-.5cm} }
\label{Phasediag}
\end{center}
\end{figure}

\emph{Asymmetric rebound --} Regardless of the advection-to-diffusion ratio, $\Pe_2$, and for sufficiently small contrast in size ($\xi-1\ll 1$), the collision always leads to a reversal of the swimming direction of both droplets (Fig.~\ref{Phasediag}, blue). The general dynamics associated with this regime is similar to that identified for symmetric collisions~\cite{Lippera19collisions}: as the droplets get closer, solute accumulates between them, reducing the polarity of their surface concentration, and effectively acting as a chemical repulsion. {This chemical asymmetry vanishes and reverses sign for a relative distance $d$ that decreases with $\Pe_2$; as a result, the reversed mechanical stresses propel the droplets away from each other.} It should be noted however that the variations of each droplet's axial velocity are not symmetric with respect to the collision. Specifically, the droplets temporarily swim away more slowly than in their final approach phase, {and this} effect is even more pronounced for the larger droplet{. For symmetric collisions, this delayed dynamic reversal of the concentration polarity was identified as the result of slower diffusion (larger $\Pe$)~\cite{Lippera19collisions}; this observation likely explains the difference of behaviour between the two droplets ($\Pe_1\geq\Pe_2$).}\\

\emph{Chasing --} For moderate $\Pe_2$ and larger size contrast $\xi$, {a chasing regime is observed where instead both droplets  eventually swim in the same direction}. While the smaller droplet still experiences a similar repulsion and rebound dynamics, the effective repulsion it exerts on the larger droplet is not sufficient to reverse its swimming direction (Fig.~\ref{Phasediag}, green). As the smaller droplet swims away, the polarity of the larger droplet still allows it to accelerate and recover its original velocity, effectively chasing the smaller droplet. {The absolute self-propulsion velocity of an isolated droplet increases with its radius (when rescaled by its radius it increases linearly with $\Pe_i$ before reaching a plateau~\cite{izri2014self,Lippera19collisions}). As a result, when far enough apart, the larger droplet will swim faster than and catch up with the smaller one. A bound state may then arise, where the two droplets maintain a fixed distance $d_\textrm{eq}$ and swim together.  Their common }velocity results from the balance of their self-propulsion properties and mutual chemical repulsion.

\begin{figure}
\begin{center}
\includegraphics[width=.6\textwidth]{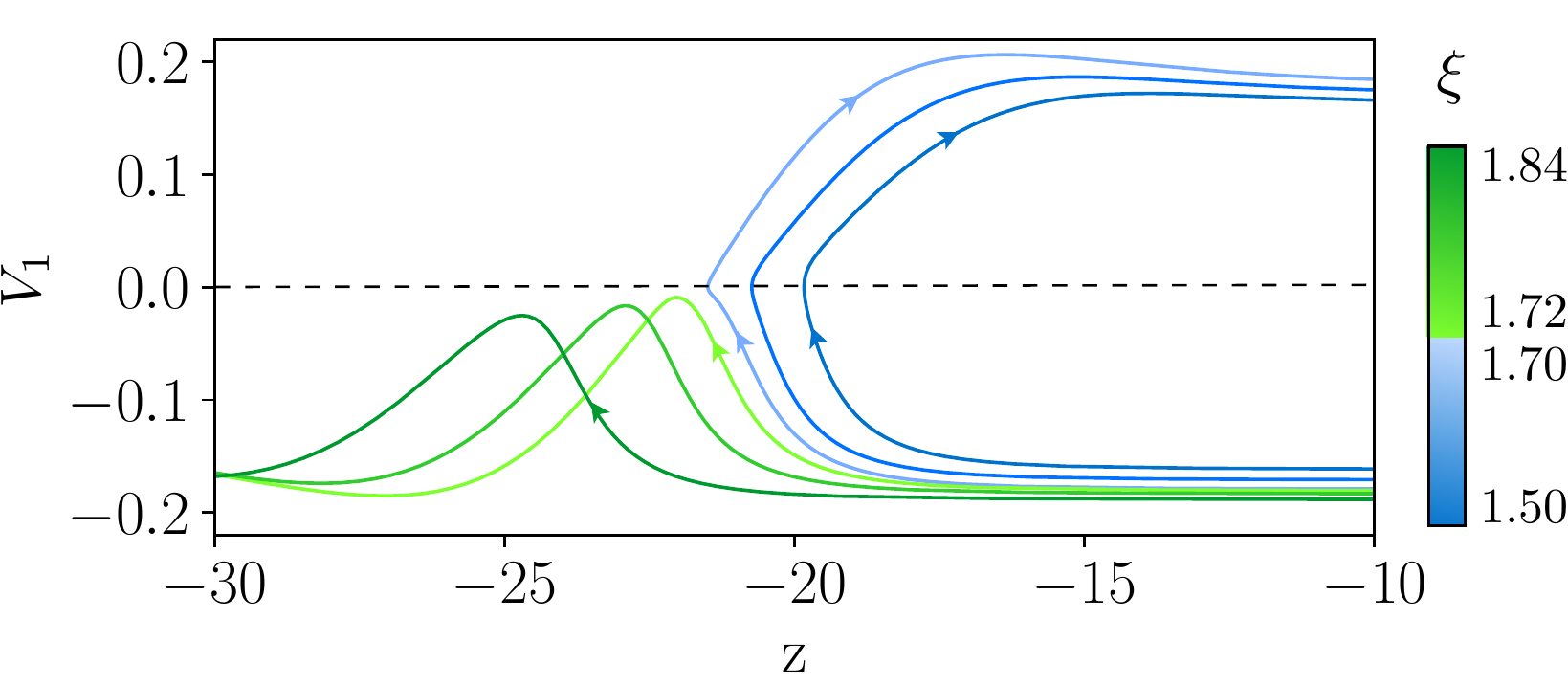}\vspace{-.3cm}
\caption{{Transition between the rebound (blue) and chasing (green) regimes}: influence of the size ratio $\xi$ on the evolution of the larger droplet velocity, $V_1$, parameterized by the droplet position. \vspace{-.6cm}}\label{transition}
\end{center}
\end{figure}

For fixed $\Pe_2$, the rebound-to-chasing transition with increasing $\xi$, illustrated in Fig.~\ref{transition}, depends on the detailed dynamics of the {collision} and thus occurs at a precise value $\xi_c(\Pe_2)$ (Figure~\ref{Phasediag}a). {This transition as $\xi$ increases results from two effects that prevent the larger droplet's rebound. The smaller droplet being exposed to an increased chemical gradient is quickly repelled away, reducing the interaction time with the larger one. Additionally, the increased size (or $\Pe$) of the larger droplet enhances the persistence of its chemical polarity and self-propulsion, as already observed for symmetric collisions~\cite{Lippera19collisions}.}

\begin{figure}
\centering
\includegraphics[width=.55\textwidth]{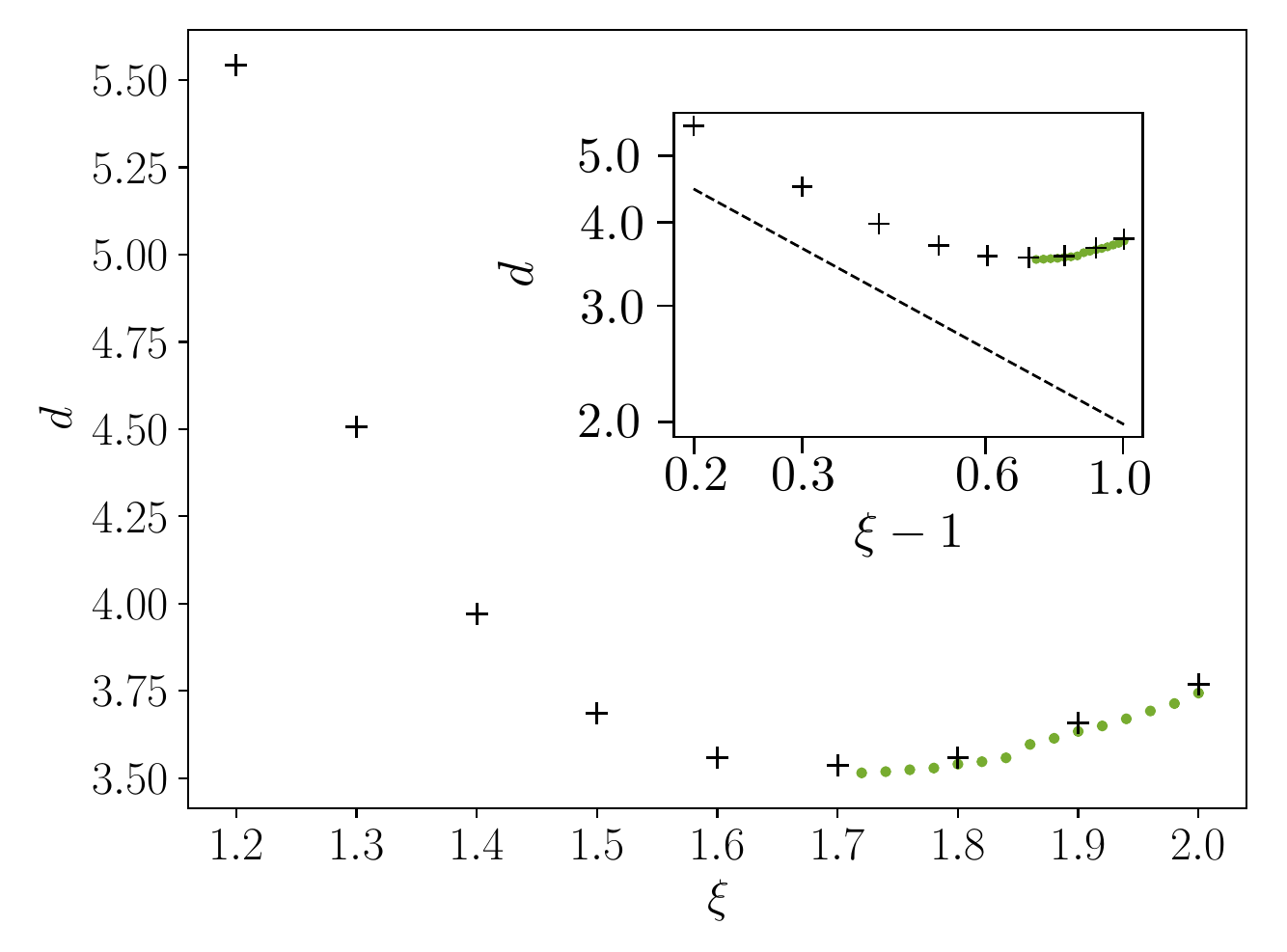}\vspace{-.5cm}
\caption{Evolution with the size ratio $\xi$ of the equilibrium distance $d_{\text{eq}}$ between the droplets in the bound state for $\Pe_2=6$. {The predictions of Eq.~\eqref{EquilibirumDistance} are also reported (dashed line in the log-log inset)}. Results are obtained from the time-dependent dynamics of two droplets initially swimming in the same direction (black crosses) or toward each other (green dots). \vspace{-.3cm}}
\label{EquilibirumDistance}
\end{figure}

The emergence of the stationary chasing regime is analysed here further. In this bound state, both droplets have the same swimming velocity $V^\infty_2<V_b<V_1^\infty$ with $V_i^\infty$ the velocity of each droplet when isolated{. Each droplet is repelled by the chemical field created by the other one: in comparison with their isolated dynamics, the larger chasing droplet is therefore slowed down while the smaller chased droplet is accelerated.}
 Increasing the distance between the droplets reduces both effects leading to the larger droplet swimming faster than, and { catching up with}, the smaller droplet. Conversely, a reduction in $d$ induces larger chemical repulsions and the resulting acceleration (resp. deceleration) of the smaller (resp. larger) droplet. {These arguments are consistent with} the existence of a stable equilibrium distance $d_\textrm{eq}$; {its} evolution with $\xi$ for fixed $\Pe_2$ is shown on Fig.~\ref{EquilibirumDistance}.

Following Ref.~\cite{Lippera19collisions}, these arguments can be formulated quantitatively in the asymptotic limit of small supercriticality, i.e. when $\delta_i=\Pe_i-\Pe_c$ are both small (which imposes $\xi-1\ll 1$). The velocity of droplet $i$ is then obtained at leading order as a function of the leading chemical gradient generated by droplet $j$,
\begin{align}
\label{Velo}
|V_i|&=\frac{V^{\infty}_i}{2}\left( 1+\sqrt{1+\textrm{sgn}(V_i)\frac{256G_{j\rightarrow i}}{\delta_i^2d^2}}   \right),
\end{align}
where $V^{\infty}_i=\delta_i/16$ denotes the reference velocity of each droplet for small $\delta_i$ and small $\xi-1$~\cite{morozov2019self}. In Eq.~\eqref{Velo}, $G_{j\rightarrow i}$ is proportional to the chemical repulsion exerted by droplet $j$ on droplet $i$  and may either slow down or speed up the motion of the latter depending on the relative signs of $V_i$ and $G_{j\rightarrow i}$. Its exact expression depends on the swimming direction of $j$ to account for the exponential (resp. algebraic) decay of concentration in front (resp. behind) that swimming droplet:
\begin{align}
\label{g1}
 G_{1\rightarrow 2}= &\left\{
      \begin{array}{ll}
        -1 & \text{if } V_1>0,\\
        - \mathrm{exp}\left[-4d|V_1|\right](1+4d|V_1|) & \text{if } V_1<0,
      \end{array}
    \right.\qquad
G_{2\rightarrow 1}=&\left\{
      \begin{array}{ll}
        1 & \quad\,\,\,\text{if } V_2<0, \\
        \mathrm{exp}\left[-4d|V_2|\right](1+4d|V_2|)& \quad\,\,\,\text{if } V_2>0.
      \end{array}
    \right.
\end{align}
{Solving Eqs.~\eqref{Velo}--\eqref{g1} for each droplet}, and considering the situation { of a leading smaller droplet}, the leading order value when $\xi-1\ll 1$ of the equilibrium distance $d_\textrm{eq}$ for which $V_1=V_2<0$ is obtained as
\begin{eqnarray}
d_{\text{eq}}=\frac{2\sqrt{2}}{\sqrt{(\xi-1)(\Pe_2-4)}}\cdot
\end{eqnarray}
This result quantifies the physical divergence of $d_\textrm{eq}$ when both droplets have the same size or swim near the critical threshold: in both cases, the swimming velocities are small so that their difference can only be compensated by weak chemical repulsion (or large inter-droplet distance).

The results of the full model, Fig.~\ref{EquilibirumDistance}, {are consistent} with the $(\xi-1)^{-1/2}$ scaling despite the large value of $\delta_2$, which is the most likely origin of any difference in prefactor.  It should be further noted from Fig.~\ref{EquilibirumDistance} that $d_\textrm{eq}$ presents a minimum for intermediate $\xi$, which is consistent with its divergence for small $\xi$. For large $\xi$, the influence of the chemical repulsion on the velocity of the larger droplet is negligeable and the whole assembly swims at a velocity {$V=V_1^\infty\gg V_2^\infty$}. {This velocity scales in dimensionless units as $\xi$, and is much larger than the spontaneous self-propulsion of the smaller droplet $V_2^\infty$. Its motion is therefore dominated by its passive drift in the chemical gradient of the larger droplet} which scales as $\xi^2/d^2$: {as a result,} $d_\textrm{eq}\sim \xi^{1/2}$ {becomes an increasing function of $\xi$}.

Figure~\ref{EquilibirumDistance} in fact reports two sets of results for the long-time dynamics of two droplets that are initially swimming in the same direction (black crosses) or toward each other (green dots). 
While the bound states emerging from this chasing regime {exist for any $\xi$}, they can not follow a collision when $\xi$ is too small ($\xi\leq 1.7$ when $\Pe_2=6$). The exact history of the droplets' motion therefore appears critical in setting their long-term dynamics, thus underlining the nonlinearity and complexity of the transition to such bound states.\\

 \begin{figure}[t!]
\begin{center}
\includegraphics[width=.33\textwidth]{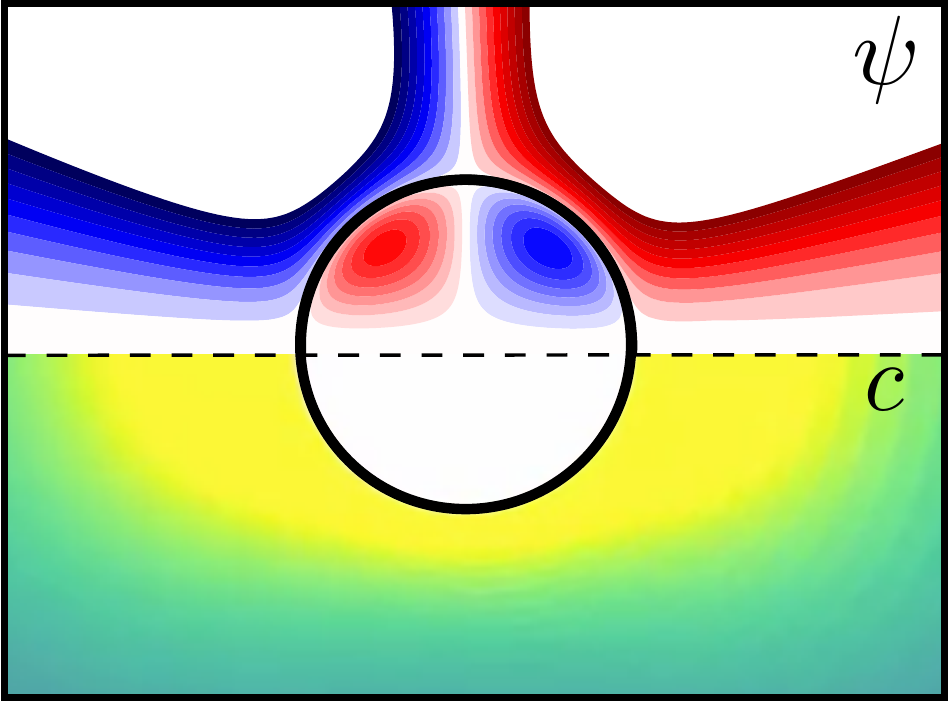}
\caption{Pausing regime: Streamlines (top) and solute concentration (bottom) around the arrested larger droplet after collision in the pausing regime obtained for $\Pe=12$ and $\xi=1.6$.\vspace{-.7cm}}
\label{stopping}
\end{center}
\end{figure}

\emph{Pausing -- } For larger $\Pe_2$, and even for relatively small contrast in size, a {pausing} regime arises. The larger droplet is slowed down by the excess concentration generated by the approach with the smaller droplet. Very much like for the other regimes, the smaller droplet quickly reverses direction and propels away from the collision site. However, the chemical repulsion experienced by the larger droplet during this brief encounter {is neither sufficiently large nor long} to reverse its chemical polarity and provoke its rebound~\cite{Lippera19collisions}; but it is sufficient to accumulate solute at its front, resulting in a symmetric but non-uniform surface concentration, driving pusher-like Marangoni flows from the equatorial plane toward its poles. Such quadrupolar flow {balances} diffusion and {maintains a}  stationary regime where the droplet acts as a symmetric pump (see Fig.~\ref{stopping}). The specific P\'eclet number of the larger droplet in this regime, ${\Pe_1=\xi^2\Pe_2}$, is greater than the instability threshold of the quadrupolar mode for a non-deformable active droplet~\cite{morozov2019self}. The emergence of the pausing regime can therefore be interpreted as the non-linear transition between two branches of the single droplet dynamics, in response to the finite perturbation induced by the collision.

The stability analysis of this pumping state lies beyond the scope of the present Letter and is left for future research. Yet, regardless of the detailed stability properties, the dominance of the dipolar propelling mode in the dynamics of a single non-deformable droplet~\cite{michelin2013spontaneous,morozov2019self,morozov2019nonlinear}, suggests that sufficiently large perturbations of the concentration or flow fields (e.g. by another droplet) will likely provoke a new mode switching and self-propulsion of the larger droplet. This may however occur long after the {first collision} so that memory of the initial propulsion direction will be lost, reminiscent of the run-and-tumble motion of swimming bacteria~\cite{berg1993,lauga2016}. This is in stark contrast with the bouncing and chasing regimes, which do preserve the collision's directionality. Such memory loss is expected to significantly affect the { long-term collective }dynamics of droplets, by introducing an effective rotational diffusion.\\

In summary, we showed in this {Letter} how variability in the size of active droplets profoundly affects their collective dynamics as a result of the sensitivity of the collision outcome to the precise droplet characteristics. Although not considered here explicitly, variability in chemical properties  likely has a similar effect, since regime selection results mainly from the droplet's chemical signature intensity and specific P\'eclet number. Once again, the strength of the advective coupling is a key factor: for moderate advection, the dynamics is only weakly modified from the symmetric collision, and large size contrast is required for more complex regimes. However, for large advective effects, non-symmetric bouncing, chasing or scattering may develop, even for droplets of comparable sizes, stressing the extreme sensitivity of the interaction.

A major limitation of the present study lies in the axisymmetric assumption, which locks the droplets' dynamics along a single axis. As a result, a $180^\circ$ flip of at least one droplet's motion is the only possible outcome of the collision, which may be avoided through small scattering in more generic situations. The stability of these axisymmetric regimes (i.e. their sensitivity to small fluctuations in the initial droplets' alignment) also remains elusive. Yet, by allowing for a complete resolution of the fully-coupled problem, this framework provides significant in-depth physical insight on the nonlinear chemo-hydrodynamic interactions of active droplets, which may prove essential in future theoretical characterization and experimental interpretation of more generic collisions, as well as for understanding the resulting scattering and collective dynamics.

\acknowledgements
This project has received funding from the European Research Council (ERC) under the European Union's Horizon 2020 research and innovation programme (grant agreement no. 714027 to S.M.).


\end{document}